\documentclass[floats,preprint,superscriptaddress,nofootinbib]{revtex4}
\usepackage{bm}
\usepackage{epsfig}
\usepackage{rotating}
 
\newcommand{\nn}{\nonumber}
\newcommand{\beq}{\begin{equation}}
\newcommand{\eeq}{\end{equation}}
\newcommand{\bea}{\begin{eqnarray}}
\newcommand{\eea}{\end{eqnarray}}
\newcommand{\ben}{\begin{eqnarray*}}
\newcommand{\een}{\end{eqnarray*}}

\def\D0{D\O}

\newcommand{\vsl}{\not\!v}

\begin{document}

\title{Even- and Odd-Parity Charmed  Meson Masses in Heavy Hadron Chiral
Perturbation Theory}%
 
\author{Thomas Mehen}%
\email{mehen@phy.duke.edu}
\affiliation{Department of Physics, Duke University, Durham NC 27708, USA}
\affiliation{Jefferson Laboratory, 12000 Jefferson Ave., Newport News VA 23606}
 
\author{Roxanne P. Springer}%
\email{rps@phy.duke.edu}
\affiliation{Department of Physics, Duke University, Durham NC 27708, USA}

\begin{abstract}
 
We derive mass formulae for the ground state, $J^P$ = $0^-$ and $1^-$, and
first excited even-parity, $J^P$ = $0^+$ and $1^+$, charmed mesons including
one loop chiral corrections and $O(1/m_c)$ counterterms in heavy
hadron chiral perturbation theory.  We show that including these counterterms is critical
for fitting the current data.  We find that certain parameter relations in the
parity doubling model are not renormalized at one loop, providing a
natural explanation for the observed equality of the hyperfine splittings of
ground state and excited doublets.

\end{abstract}
 
\date{\today}
\maketitle

\section{Introduction}

Excited charmed mesons with angular momentum and parity $J^P = 0^+$
and $1^+$ have been observed in several experiments. The
masses of the $J^P = 0^+$ and $1^+$ charmed strange mesons,
$D_s(2317)$ and $D_s(2460)$~\cite{Aubert:2003fg,Besson:2003jp}, are
below threshold for decays into ground state charmed mesons and
kaons. The only strong decay modes are via isospin-violating $\pi^0$
emission, making the states quite narrow ($\Gamma < 5.5$ MeV).  Other
experiments~\cite{Anderson:1999wn,Abe:2003zm,Link:2003bd} claim to
observe the nonstrange $J^P=0^+$ and $1^+$ states. These states can
decay to the ground states by $S$-wave pion emission and therefore are
quite broad ($\Gamma \sim 300$ MeV).

The spectrum of the $J^P=0^+$ and $1^+$ charmed mesons presents a
number of puzzles for theory. Before their discovery, quark model and
lattice calculations predicted that the masses of the $J^P = 0^+$ and
$1^+$ charmed strange mesons would be significantly higher than
observed~\cite{Godfrey:xj,Godfrey:wj,Hein:2000qu,Boyle:1997rk,Lewis:2000sv}.
Further, the hyperfine splittings of all ground state charmed mesons
and the hyperfine splitting of the $D_s(2317)$ and $D_s(2460)$ are all
equal to within 2\%.  This is surprising because there is no obvious
symmetry of quantum chromodynamics (QCD) which predicts these
equalities. Finally, the $SU(3)$ splittings of the $J^P=0^+$ and $1^+$
charmed mesons are much smaller than theoretical expectations.

In the heavy quark limit, the coupling of the heavy quark spin to the light degrees of freedom in the heavy meson vanishes and the angular
momentum and parity of the light degrees of freedom, $j^p$, can be used to classify heavy meson states.  The spectrum consists of degenerate
heavy meson doublets with definite $j^p$. The $J^P = 0^-$ and $1^-$ heavy mesons are members of the $j^p=\frac{1}{2}^-$ ground state doublet.
The lowest lying excited states, the  $J^P = 0^+$ and $1^+$ heavy mesons, are members of the  $j^p=\frac{1}{2}^+$ doublet. There is also an
excited doublet of heavy mesons with $j^p=\frac{3}{2}^+$, whose members have $J^P =1^+$ and $2^+$. The 
$j^p=\frac{3}{2}^+$ mesons decay to the ground state by
$D$-wave pion emission, typically have widths $\Gamma \sim 20$ MeV, and therefore have well-measured masses.  The hyperfine splittings for all
of these heavy quark doublets are suppressed by $1/m_Q$, where $m_Q$ is the heavy quark mass. 

The experimental  data on the masses of the known charmed mesons is summarized in Table~\ref{mesondata}.  The lowest lying flavor $SU(3)$ 
anti-triplets are $J^P=0^-$ ($c \overline u$, $c \overline d$, $c \overline s$) $\sim$ ($D^0$, $D^+$, $D_s^+$) and $J^P=1^-$ ($D^{*0}$,
$D^{*+}$,$D_s^{*+}$). The first excited states are $J^P=0^+$ ($D_0^0$, $D_0^+$,$D_{0s}^+$) and $J^P=1^+$ ($D_1^{0 \prime}$, $D_1^{+ \prime}$, $D_{1s}^{+
\prime}$). The members of the $j^p=\frac{3}{2}^+$ doublet are $J^P = 1^+$ ($D_1^{0}$, $D_1^{+}$, $D_{1s}^{+}$) and $J^P = 2^+$ ($D_2^{0}$, $D_2^{+}$,
$D_{2s}^{+}$).  Not shown is a narrow charmed strange meson, $D^+_s(2632)$, recently observed by the SELEX collaboration~\cite{Evdokimov:2004iy}. The 
spin and parity of this meson and its place in the  charmed meson spectrum is currently unknown.
\begin{table}[t!] 
  \begin{tabular}{cccccccc}
     $j^p$&$J^P$&$c\bar{u}$&& $c\bar{d}$ && $c\bar{s}$ &\\
     \hline\hline
    & &name&$M ({\rm MeV})$&name& $M ({\rm MeV})$&name&$M ({\rm MeV})$  \\
$3/2^+$ & $2^+$&$D^0_2$ & $2458.9 \pm 2.0$& $D^+_2$ &  $2459 \pm 4$ & $ D^+_{s2}$ 
    & $2572.4 \pm 1.5$ \\
     \hline
$3/2^+$ & $1^+$&$D^0_1$ & $2422.2 \pm 1.8$& $D^+_1$ &  $2427 \pm 5$ & $ D^+_{s1}$ 
    & $2535.4  \pm 0.6$ \\
     \hline
    $1/2^+$& $1^+$&$D^{0 \prime}_1$ & $2438 \pm 31$& - &  - & $ D^{+ \prime}_{s1}$ & $2459.3 \pm 1.3$ \\
     \hline
     $1/2^+$&$0^+$&$D^0_0$ & $2308 \pm 36$& - &  - & $ D^+_{s0}$ & $2317.4 \pm 0.9$ \\
     \hline
     $1/2^-$&$1^-$&$D^{*0} $ & $2006.7 \pm 0.5$& $D^{*+}$  &  $2010.0 \pm 0.5$ & 
    $ D_s^{*+}$ & $2112.1 \pm 0.7$ \\
     \hline
     $1/2^-$&$0^-$&$D^0 $ & $1864.6 \pm 0.5$&  $D^{+}$ &  $1869.4 \pm 0.5$  & 
    $ D_s^+$ & $1968.3 \pm 0.5$ \\
     \hline\hline
  \end{tabular}
  \caption{The spectrum of charmed mesons. $j^P$ is the angular momentum and parity of the light degrees of freedom. 
$J^P$ is the angular momentum and parity of the meson.}
  \label{mesondata}
\end{table}
For all mesons except the nonstrange $j^p=\frac{1}{2}^+$ doublet, we use numbers from the Particle Data Group~\cite{Eidelman:2004wy}.
For nonstrange $j^p=\frac{1}{2}^+$  mesons,  we use the Belle~\cite{Abe:2003zm} measurement of the $D_0^0$ mass and average the  CLEO~\cite{Anderson:1999wn}
and Belle~\cite{Abe:2003zm} measurements of the $D_1^0$ mass. \footnote{
The FOCUS collaboration reports structures in excess of background in the $D^+\pi^-$
and $D^0\pi^+$ invariant mass spectra which  could be interpreted as  scalar resonances \cite{Link:2003bd}. However, if these resonances exist their
masses are 99 MeV higher than the Belle measurement and 80 MeV higher than the mass of the $D_{s0}^{+}$. It is implausible that such resonances are
related to the $D_{s 0}^+$ by $SU(3)$ symmetry so we do not use this data to determine the $J^P=0^+$ nonstrange meson masses.}

As stated earlier, the hyperfine splittings of the $j^p=\frac{1}{2}^-$ and $j^p=\frac{1}{2}^+$ doublets are nearly equal. The
known hyperfine splittings of the $j^p=\frac{1}{2}^-$ and $j^p=\frac{1}{2}^+$ charmed mesons are:
\bea\label{hyperfine}
m_{D^{*0}}-m_{D^0} &=& 142.1 \pm 0.07 \,{\rm MeV} \nonumber \\
m_{D^{*+}}-m_{D^+}&=& 140.6 \pm 0.1 \,{\rm MeV} \nonumber \\
m_{D_s^{*+}}-m_{D_s^+} &=& 143.8 \pm 0.4 \,{\rm MeV} \nonumber \\
m_{D_{s1}^{+\prime}}-m_{D_{s0}^+} &=& 141.9  \pm 1.6 \,{\rm MeV} \nonumber \\
m_{D_{1}^{0\prime}}-m_{D_{0}^0} &=& 130 \pm 48 \,{\rm MeV} \, .
\eea
Here, the first three numbers are the hyperfine splittings quoted by the Particle Data Group~\cite{Eidelman:2004wy}. The last two numbers  are
obtained by taking the difference of the masses in Table \ref{mesondata}. The error in the  last two lines  of Eq.~(\ref{hyperfine}) is obtained  
by adding the errors in the individual masses in quadrature.   All four hyperfine splittings which
have been measured accurately are $\approx 142$ MeV to within 2 MeV or less. Hyperfine splittings in different  heavy quark doublets are unrelated
by heavy quark symmetry. For example, the hyperfine splitting for  $j^p=\frac{3}{2}^+$ doublets is $\sim 40$ MeV, which differs significantly from
the  $j^p=\frac{1}{2}^-$ and $\frac{1}{2}^+$ hyperfine splittings. In the $SU(3)$ limit, the hyperfine splittings of nonstrange and strange ground
state mesons are the same. That this $SU(3)$ prediction holds to within  2\%  is surprising
given the typical size of $SU(3)$ breaking
effects in QCD. 

Another  puzzling feature of the spectrum is the pattern of $SU(3)$ violation in the splittings within the 
even-parity doublets. Finite light quark
($m_u$, $m_d$, and $m_s$) masses and electromagnetic effects cause flavor-splitting among the mesons.  The isospin splitting seen in the charmed meson mass
spectrum is of expected size, but the splitting between the strange and nonstrange sector is unexpected.  The  mass difference between strange and nonstrange
mesons whose other quantum numbers are identical  is expected to be $\sim 100$ MeV.  For the ground state charmed mesons this is indeed the case.  For the excited
states, however, the $SU(3)$ breaking is 
\bea 
m_{D_{s1}^{+\prime}} - m_{D_{1}^{0 \prime}} &=& 21 \pm 31 \,{\rm MeV} \nn \\
m_{D_{s0}^{+}} - m_{D_{0}^{0}} &=& 9 \pm 36 \,{\rm MeV} \,.
\eea
Even allowing for the large errors due to  the uncertainty in the masses of the nonstrange $j^p=\frac{1}{2}^+$ charmed mesons, the $SU(3)$ splitting
is far below theoretical expectations. 

The $D_s(2317)$ and $D_s(2460)$ are only 40 MeV below the $D K$ and $D^* K$ threshold, respectively. This fact as well as the puzzles mentioned above have  led to the
hypothesis that they are  bound states of $D^{(*)}$ and $K$~\cite{Barnes:2003dj,Nussinov:2003uj,Chen:2004dy}.  Several papers analyze the spectroscopy of excited charm
mesons by extending the quark model to include couplings to the $D K$ continuum. This coupled channel effect has been analyzed within the  quark model
\cite{Hwang:2004cd}, chiral quark models \cite{Lee:2004gt,Simonov:2004ar} as well as unitarized meson
models~\cite{vanBeveren:2003kd,vanBeveren:2003jv,vanBeveren:2004ve}. The unitarized meson model has also been used to make predictions for the spectroscopy of excited
$B$ mesons~\cite{vanBeveren:2003jv,vanBeveren:2004bz}. However, if one assumes that the  $D_s(2317)$ and $D_s(2460)$ are nonrelativistic $D K$ and $D^* K$ bound
states, respectively, heavy-hadron chiral perturbation theory (HH$\chi$PT)~\cite{hhchpt} can be used to predict their electromagnetic branching ratios. These predictions
are found to be in serious disagreement with experiment~\cite{Mehen:2004uj}. On the other hand, if one assumes that the  $D_s(2317)$ and $D_s(2460)$ are conventional
states, then HH$\chi$PT predictions for strong and electromagnetic decays are consistent with available data~\cite{Mehen:2004uj,Colangelo:2003vg}. An alternative
interpretation of these particles as exotic $c\bar s\bar q q$ tetraquarks has also been
proposed~\cite{Nussinov:2003uj,Cheng:2003kg,Terasaki:2003qa,Terasaki:2003dj,Terasaki:2004yx,Vijande:2003wk}. For a review of theoretical work on $D_s(2317)$ and
$D_s(2460)$, see Ref.~\cite{Colangelo:2004vu}.

In this paper we analyze the spectroscopy of charmed mesons using HH$\chi$PT. This theory can be used to analyze the low energy strong interactions of heavy mesons in a
systematic expansion  in  light quark masses, $m_q$, and inverse heavy quark masses, $1/m_Q$. Nonanalytic corrections from loops with Goldstone bosons can be
calculated in this formalism.  The masses of the ground state heavy mesons have been studied in the heavy quark limit~\cite{Isgur:1989vq,Isgur:1989ed}, including
leading corrections from finite heavy quark masses and nonzero light quark
masses~\cite{Rosner:1992qw,Randall:1992pb,Jenkins:1992hx,DiBartolomeo:1994ir,Blok:1996iz,yeh_lee}. These papers use a version of HH$\chi$PT  which includes only the
lowest lying $j^p = \frac{1}{2}^-$ heavy quark doublets.   Many recent studies of excited $J^P=0^+$ and $1^+$ heavy mesons use Lagrangians that include only
$j^p=\frac{1}{2}^-$ and $j^p=\frac{1}{2}^+$ heavy quark doublets as explicit degrees of freedom. However, the excited $j^p=\frac{3}{2}^+$ doublets are only separated
from the $j^p=\frac{1}{2}^+$ doublets by $\lesssim$ 130 MeV. Further, the $j^p =\frac{3}{2}^+$ doublets couple to the  $j^p=\frac{1}{2}^+$ doublets at leading order in
the chiral expansion, while the coupling of the $j^p=\frac{3}{2}^+$ doublets to the ground state doublets is higher order in the chiral expansion~\cite{Falk:1992cx}.
For these reasons, loops with virtual excited $j^p=\frac{3}{2}^+$ could have important effects on the physics of $j^p=\frac{1}{2}^+$ doublets. In this paper we will
study the version of HH$\chi$PT containing only the $j^p=\frac{1}{2}^-$ and $j^p=\frac{1}{2}^+$  heavy quark doublets and leave investigation of loop effects from more
highly excited states for future work.

A model of heavy mesons closely related to HH$\chi$PT is the parity
doubling model of
Refs.~\cite{Bardeen:1993ae,Bardeen:2003kt,Nowak:1992um,Nowak:2003ra}. The
parity doubling model is the analog of the linear sigma model for
heavy mesons.  Heavy meson doublets transforming linearly under
$SU(3)_L \times SU(3)_R$ couple in a chirally invariant way to a field
$\Sigma$ transforming in the $(\bar 3, 3)$ of $SU(3)_L \times
SU(3)_R$. The field $\Sigma$ develops a vacuum expectation value and
the resulting theory of heavy mesons has the same form as HH$\chi$PT
for the low lying odd- and even-parity doublets. Unlike HH$\chi$PT,
the parity doubling model predicts relationships among otherwise
independent parameters in the theory. One important prediction is that
the hyperfine splittings of the $j^p=\frac{1}{2}^-$ and $j^p
=\frac{1}{2}^+$ doublets are equal at tree level. This interesting
prediction could partially explain the observed pattern of heavy meson
hyperfine splittings, but it is not clear from
Refs.~\cite{Bardeen:1993ae,Bardeen:2003kt,Nowak:1992um,Nowak:2003ra}
whether this prediction survives beyond tree level. This is a concern
because loop corrections in HH$\chi$PT can be significant.

In this paper, we calculate the one loop HH$\chi$PT corrections to the
masses of $j^p = \frac{1}{2}^-$ and $j^p =\frac{1}{2}^+$ heavy meson
doublets.  We include all $O(1/m_Q)$ heavy quark spin symmetry
violating operators that appear to this order. A brief review of the
HH$\chi$PT formalism is given in section II and explicit formulae for
the masses at one loop appear in the Appendix.  In section III, we
attempt to fit the observed mass spectrum with our one-loop
formulae. The large number of free parameters makes it possible to reproduce  
the spectrum of $j^p=\frac{1}{2}^-$ and $j^p=\frac{1}{2}^+$ charmed mesons.
In the $m_Q \to \infty$ limit our
calculation of the difference of the $SU(3)$ splittings in HH$\chi$PT
agrees with Ref.~\cite{Becirevic:2004uv}. Our analysis differs from
that in Ref.~\cite{Becirevic:2004uv} in that we include $1/m_Q$ operators
and perform a global fit to the
spectrum with all counterterms treated as free parameters.  In the
approximation used in Ref.~\cite{Becirevic:2004uv} there is a single 
counterterm constrained using lattice data.

In section IV, we examine corrections to the hyperfine splittings and discuss the naturalness of the parity doubling model. The parity doubling model
predicts that the hyperfine splittings and the magnitudes of the axial couplings of the $j^p=\frac{1}{2}^-$ and $j^p = \frac{1}{2}^+$ doublets are equal
at tree level. We find that these parameter relations are preserved by the one loop corrections so that the model provides a natural explanation for the equality of hyperfine
splittings.  Finally, in section V, we use heavy quark effective theory (HQET) to estimate the masses of  the $j^p=\frac{1}{2}^+$ $B$ mesons, which have not 
yet been observed. These predictions may be helpful to experimentalists looking for these states.

\section{HH$\chi$PT Mass Counterterms}

In HH$\chi$PT, the ground state doublet is represented by the fields~\cite{hhchpt}
\begin{eqnarray}\label{H}
H_a = \frac{1\, +\!\vsl}{2}\left(H^\mu_a \gamma_\mu - H_a \gamma_5 \right) \, ,
\end{eqnarray}
where $a$ is an $SU(3)$ index. In the charm sector, $H_a$ consists of the $(D^0,D^+,D_s^+)$  pseudoscalar mesons
and  $H^\mu_a$ are the $(D^{*0},D^{*+},D_s^{*+})$ vector mesons. The $j^p=\frac{1}{2}^+$ 
doublet is represented by the fields~\cite{Falk:1991nq}
\begin{eqnarray}\label{S}
S_a = \frac{1\, +\!\vsl}{2}\left(S^\mu_a \gamma_\mu \gamma_5 - S_a \right) \, ,
\end{eqnarray}
where the scalar states in the charm sector are $S_a = D_{0a}$ and the axial vectors are $S^\mu_a =D_{1 a}^{ \prime}$.  
The kinetic terms of these fields are included in:
\begin{eqnarray}
{\cal L}^{kinetic}_v &=&  - {\rm Tr} [\overline H_a (i v \cdot D_{ba}- 
\delta_{H}\,\delta_{ab}) H_b ] + {\rm Tr}[\overline S_a ( i v \cdot D_{ba} - 
\delta_{S}\, \delta_{ab}) S_b]   \, ,
\end{eqnarray}
where $\delta_{H}$ and  $\delta_S$   are the residual masses of the $H$ and $S$  fields, respectively, and $D_{ba}$ is the chirally covariant derivative. In the theory with
only $H$ fields one is free to  set $\delta_H =0$. Since the only dimensionful parameters entering the loops in this theory  are hyperfine splittings and
meson masses, the UV divergences (in dimensional regularization) vanish in the $m_q \to 0$ and $m_Q \to \infty$ limit. Divergences in loop corrections are
canceled by counterterms which are $O(m_q)$ or $O(1/m_Q)$. Once the  $S$  fields are added to the theory, there is another dimensionful quantity, $\delta_S
- \delta_H$, which does not vanish  as $m_q \to 0$ and $m_Q \to \infty$. The $H$ self-energy diagrams with virtual $S$ fields give a UV divergent
contribution which survives in the $m_q \to 0$ and $m_Q \to \infty$ limit. Such a divergence must be canceled by a mass counterterm which respects $SU(3)$
and heavy-quark spin symmetry and the only available counterterm is $\delta_H {\rm Tr}{\overline H_a} H_a$. However,  after one-loop divergences are
canceled one is free to define the finite part of $\delta_H$ for convenience.  

The fields have axial couplings to the pseudo--Goldstone bosons,
\begin{eqnarray}\label{axial}
{\cal L}^{axial}_v &=& g \, {\rm Tr} [ \overline H_a H_b \, \slash \hskip -.25cm A_{ba}
\gamma_5] + g^\prime \, {\rm Tr} [ \overline S_a S_b \, \slash \hskip -.25cm A_{ba}
\gamma_5] + h \, {\rm Tr} [ \overline H_a S_b \, \slash \hskip -.25cm A_{ba} \gamma_5
+ {\rm h.c.}]   \, ,
\end{eqnarray}
where $g$, $g^\prime$, and $h$ are dimensionless constants to be determined from experiment. The other terms in the Lagrangian 
required are higher order mass counterterms.  We use the notation of 
Ref.~\cite{Jenkins:1992hx} and generalize it to include
the $S$ fields as well as the $H$ fields. 
\begin{eqnarray}\label{mass}
{\cal L}_v^{\rm mass} &=& -{\Delta_H \over 8} {\rm Tr}
[\overline H_a \sigma^{\mu \nu} H_a \sigma_{\mu \nu}] +
{\Delta_S \over 8} {\rm Tr}[\overline S_a \sigma^{\mu \nu} S_a \sigma_{\mu \nu}] \, \nn \\
&&+ a_H {\rm Tr} [\overline H_a H_b]\,m^\xi_{ba} - a_S {\rm Tr} [\overline S_a S_b] \, m^\xi_{ba} + 
\sigma_H {\rm Tr} [\overline H_a H_a] \, m^\xi_{bb} -
\sigma_S {\rm Tr} [\overline S_a S_a] \, m^\xi_{bb}\, \nn \\
&& 
-{\Delta_H^{(a)} \over 8} {\rm Tr} [\overline H_a \sigma^{\mu \nu} H_b
\sigma_{\mu \nu}] \, m^\xi_{ba}+ 
{\Delta_S^{(a)} \over 8} {\rm Tr} [\overline 
S_a \sigma^{\mu \nu} S_b \sigma_{\mu \nu} ] \, m^\xi_{ba}  \nonumber \\
&& - {\Delta_H^{(\sigma)} \over 8} 
    {\rm Tr} [\overline H_a \sigma^{\mu \nu} H_a \sigma_{\mu \nu}]\, m^\xi_{bb} +
{\Delta_S^{(\sigma)} \over 8} 
    {\rm Tr} [\overline S_a \sigma^{\mu \nu} S_a \sigma_{\mu \nu}]\, m^\xi_{bb} \, ,
\end{eqnarray}
where $m^\xi_{ba} = \frac{1}{2}(\xi m_q \xi + \xi^\dagger m_q \xi^\dagger)_{ba}$,  $m_q = {\rm diag}(m_u, m_d, m_s)$
and $\xi = \sqrt{\Sigma} = \exp(i \Pi/f)$, where $\Pi$ is the matrix of Goldstone bosons.  The first
line in Eq.~(\ref{mass}) contains spin-symmetry violating operators which give rise to hyperfine splittings. 
The second line contains terms which are $O(m_q)$ and preserve  heavy-quark spin
symmetry. Finally, the last two lines contain terms which are $O(m_q)$ and violate heavy-quark spin symmetry. 

\begin{figure}[!t]
  \centerline{\epsfysize=5.5truecm \hspace{0.4 in} \epsfbox[110 460 480 650]{se.ps}  }
 {\tighten
\caption[1]{One-loop self energy diagrams for the  $H$ and $S$ fields.
$H$ fields are single lines, $S$ fields are double lines and Goldstone bosons are dashed lines.}
\label{se} }
\end{figure}

At tree level the residual masses are 
\begin{eqnarray}\label{tree}
m^0_{H_a} &=&  \delta_H - \frac{3}{4}\Delta_H + \sigma_H \, \overline{m} + a_H \,m_a 
-\frac{3}{4}\Delta_H^{(\sigma)} \, \overline{m} - \frac{3}{4}\Delta_H^{(a)} \, m_a  \nn \\
m^0_{H_a^*} &=&  \delta_H + \frac{1}{4}\Delta_H + \sigma_H \, \overline{m} + a_H \, m_a 
+\frac{1}{4}\Delta_H^{(\sigma)} \, \overline{m} + \frac{1}{4}\Delta_H^{(a)} \, m_a  \nn \\
m^0_{S_a} &=&  \delta_S - \frac{3}{4}\Delta_S + \sigma_S \, \overline{m} + a_S \, m_a 
-\frac{3}{4}\Delta_S^{(\sigma)} \, \overline{m} - \frac{3}{4}\Delta_S^{(a)} \,m_a  \nn \\
m^0_{S_a^*} &=&  \delta_S + \frac{1}{4}\Delta_S + \sigma_S \, \overline{m} + a_S \, m_a 
+\frac{1}{4}\Delta_S^{(\sigma)} \, \overline{m} + \frac{1}{4}\Delta_S^{(a)} \, m_a  
\end{eqnarray}
where $m_a= (m_u, m_d, m_s)$ and $\overline{m} =m_u+m_d+m_s$.  Here
the asterisk denotes the spin-1 member of the heavy quark doublet.  In
the isospin limit $m_u = m_d$. HH$\chi$PT is a double expansion in
$\Lambda_{\rm QCD}/m_Q$ and $Q/\Lambda_\chi$, where $Q \sim m_\pi,
m_K, m_\eta$ and $\Lambda_\chi = 4 \pi f \approx 1.5$ GeV.  The
parameters $\delta_{H}$, $\delta_{S}$, $\Delta_H$, and $\Delta_S$ are
treated as $O(Q)$ in the power counting~\cite{Mehen:2004uj}.  Since
$m_q \propto m_\pi^2 \sim Q^2$ the remaining terms in Eq.~(\ref{tree})
are formally higher order in the power counting. The loop corrections
to the masses are shown in Fig.~\ref{se}. Single lines represent the
$H$ fields and double lines represent the $S$ fields. All diagrams are
$O(Q^3)$. The loop corrections are regulated using dimensional
regularization. Complete one loop expressions for the masses are given
in the Appendix.

\section{Charmed Meson Spectrum}

In this section we analyze the charmed meson spectrum using the
one-loop mass formulae given in the Appendix.  We will work in the isospin
limit, where the masses of $H_{1}$ and $H_{2}$, for instance, are
identical.  Then there are eight different residual masses: $m_{H_1}$,
$m_{H_3}$, $m_{H_1^*}$, $m_{H_3^*}$, $m_{S_1}$, $m_{S_3}$, $m_{S_1^*}$, and
$m_{S_3^*}$.  To determine the experimental values of $m_{H_1}$ and
$m_{H_1^*}$, we average the masses of the two known isospin
states. The residual masses are defined to be the difference
between the real masses and an arbitrarily chosen reference mass of
$O(m_Q)$. We will measure all masses relative to the nonstrange spin
averaged $H$ mass, so $(m_{H_1}+3 m_{H^*_1})/4 = 0$. Therefore, the
experimentally measured residual masses we will fit to are:
\bea\label{rm} 
m_{H_1} = -106.1 \,{\rm MeV} \qquad m_{H_3} = -4.75 \,{\rm MeV}
\qquad m_{H_1^*} = 35.4 \,{\rm MeV} \qquad m_{H_3^*} = 139.1 \,{\rm MeV} \nn \\
m_{S_1} = 335.0 \,{\rm MeV} \qquad m_{S_3} = 344.4 \,{\rm MeV}
\qquad m_{S_1^*} =465.0 \,{\rm MeV}  \qquad m_{S_3^*}
= 486.3 \,{\rm MeV} \, . \eea
The tree level expressions in Eq.~(\ref{tree}) reproduce these values
with $\delta_S + \sigma_S \, \overline{m} - \delta_H - \sigma_H \,
\overline{m} = 432 \pm 26$ MeV, $\Delta_H +\Delta^{(\sigma)}_H 
\, \overline{m}= 146
\pm 1$ MeV, $\Delta_S +\Delta^{(\sigma)}_S \, \overline{m} = 129 \pm
50$ MeV, $a_H=1.14 \pm 0.06$, $a_S=0.21 \pm 0.29$, $\Delta^{(a)}_H =
-0.03 \pm 0.01$, and $\Delta^{(a)}_S = 0.14 \pm 0.55$.  The errors
used to obtain this fit are the experimental ones, dominated by
the uncertainty in the nonstrange $0^+$ and $1^+$ masses.  This gives
rise to the large uncertainties seen in parameters in the Lagrangian
involving the $S$ fields.  The
fits presented in this section use Mathematica \cite{math} and/or 
Minuit.\cite{minuit}

The loop corrections depend on eleven parameters: $g$, $g^\prime$,
$h$, $a_H$, $a_S$, $\Delta^{(a)}_H$, $\Delta^{(a)}_S$, $\delta_H + \sigma_H \,
\overline{m}$, $\delta_S + \sigma_S \, \overline{m}$, $\Delta_H
+\Delta^{(\sigma)}_H \, \overline{m}$, and $\Delta_S
+\Delta^{(\sigma)}_S \,\overline{m}$.  The parameters
$\sigma_H$, $\sigma_S$, $\Delta_H^{(\sigma)}$, and $\Delta_S^{(\sigma)}$
cannot be separately determined because they always appear in linear
combination with the parameters $\delta_H$, $\delta_S$, $\Delta_H$, and
$\Delta_S$, respectively.  Below we will absorb the contribution of the
parameters
$\sigma_H$, $\sigma_S$, $\Delta_H^{(\sigma)}$, and $\Delta_S^{(\sigma)}$
into the measured values of $\delta_H$,
$\delta_S$, $\Delta_H$, and $\Delta_S$, respectively.  

An analysis of $D^*$ decays using a one-loop calculation without explicit excited states
yields  $g = 0.27^{+0.06}_{-0.03}$~\cite{Stewart:1998ke}.
From the widths of the nonstrange resonances observed by Belle we have
extracted $h = 0.69 \pm 0.09$ at tree level~\cite{Mehen:2004uj}.  Both couplings are
of order unity and therefore consistent with naive power counting. The
remaining parameters are unknown.

We use $f = 120$ MeV, which is the value extracted in
Ref.~\cite{Stewart:1998ke} using the one loop formulae for pion and
kaon decay constants, first derived in Ref.~\cite{Gasser:1984gg}. We
set $m_u = m_d = 4$ MeV and $m_s = 90$ MeV. Below we show several
different fits.  In the first case we fix $g$ and $h$ to the values
(given above) extracted from previous analyses.  This leaves nine
remaining free parameters.  Performing a least chi-squared fit to the
meson spectrum, using experimental uncertainties, we obtain the
following central values
\bea 
m_{H_1} = -106 \,{\rm MeV} \qquad m_{H_3} = -5
\,{\rm MeV} \qquad m_{H_1^*} = 35 \,{\rm MeV} \qquad m_{H_3^*} = 139
\,{\rm MeV} \nn \\ m_{S_1} = 160 \,{\rm MeV} \qquad m_{S_3} = 344
\,{\rm MeV} \qquad m_{S_1^*} = 296 \,{\rm MeV} \qquad m_{S_3^*} = 486
\,{\rm MeV} \,. 
\eea 
The parameters extracted from this fit are:
$g^\prime =0.09 \pm$ 0.03, $\delta_H = -83 \pm 3$ MeV, $\delta_S = 244
\pm 1$ MeV, $\Delta_H = 133 \pm 2$ MeV, $\Delta_S = 136 \pm 1$ MeV,
$a_H = 1.70 \pm 0.01$, $a_S =0.25 \pm 0.08$, $\Delta_H^{(a)}= -0.07
\pm 0.01$, and $\Delta_S^{(a)}= 0.04 \pm 0.03$. Six of the mass
parameters are reproduced quite well while $m_{S_1}$ and $m_{S_1^*}$
are lower than the central values of experiments by about 175 and 169
MeV, respectively.  This qualitative picture persists without much
sensitivity to the value of $g^\prime$. However, these fits are not
very good and such a procedure may not be very realistic.  The values
of $g$ and $h$ used above were extracted using a fit to a one-loop
calculation not including the $S$ fields, and a tree-level fit,
respectively.  There is no reason to believe that these values are the
ones which are appropriate for a calculation that includes graphs with
internal $S$ states.  Note that large changes between tree- versus
loop-extracted parameter values do not necessarily indicate poor
convergence; what is important is that the observables do not suffer
large changes between orders.

If we include $g$ and $h$ as free parameters in an 11-parameter fit,
there are many solutions which yield central values identical to
the experimental residual masses given in Eq.~(\ref{rm}).  
In addition to the experimental
errors we also include 20\% ``theoretical'' errors to mimic
the fact that our analysis is only accurate to ${\cal O}(Q^3)$. 
The masses obtained are then accompanied by errors at the 
$\pm$ 30 to 40 MeV level. Examples of parameter
sets which give these results are:
\begin{itemize}
\item[(a)] $|g|=1.15 \pm 0.06$, $|g^\prime|=0.90 \pm 0.06$, $|h|=2.3 \pm 0.2$,
$\delta_H = 195 \pm 41$ MeV, $\delta_S = 332 \pm 31$ MeV,
$\Delta_H = 465 \pm 24$ MeV, 
$\Delta_S = 597 \pm 28$ MeV,
$a_H = 7 \pm 1$, $a_S =-4 \pm 1$, $\Delta_H^{(a)}= -4.4 \pm 0.7$, and
$\Delta_S^{(a)}= - 10 \pm 2$.

\item[(b)]  $|g|=0.65 \pm 0.06$, $|g^\prime|=0.89 \pm 0.08$, $|h|=0.2 \pm 0.1$,
$\delta_H = 117 \pm 21$ MeV, $\delta_S = 646 \pm 40$ MeV,
$\Delta_H = 68 \pm 42$ MeV, 
$\Delta_S = 447 \pm 23$ MeV,
$a_H = 3.8 \pm 0.7$, $a_S = 3.1 \pm 0.7$, $\Delta_H^{(a)}= -0.3 \pm 1$, and
$\Delta_S^{(a)}= - 2.8 \pm 1$.

\item[(c)]  $|g|=0.89 \pm 0.07$, $|g^\prime|=0.24 \pm 0.13$, $|h|=0.98 \pm 0.11$,
$\delta_H = 203 \pm 39$ MeV, $\delta_S = 399 \pm 26$ MeV,
$\Delta_H = 242 \pm 25$ MeV, 
$\Delta_S = 116 \pm 59$ MeV,
$a_H = 5.8 \pm 1.1$, $a_S =-1.4 \pm 1.5$, $\Delta_H^{(a)}= -1.7 \pm 0.8$, and
$\Delta_S^{(a)}= 2.1 \pm 1.7$.

\item[(d)]  $|g|=|g^\prime|=0.70 \pm 0.03$, $|h|=2.4 \pm 0.2$,
$\delta_H = 114 \pm 64$ MeV, $\delta_S = 231 \pm 36$ MeV,
$\Delta_H = 682 \pm 4$ MeV, 
$a_H = 4.3 \pm 0.7$, $a_S =-3.0 \pm 2.1$, $\Delta_H^{(a)}= -0.89 \pm 0.96$, and
$\Delta_S^{(a)}= -2.7 \pm 0.9$.  
(In this fit, the constraint $\Delta_S = \Delta_H$ + 30 MeV was used.)

\end{itemize}

There are clearly many local minima which Minuit \cite{minuit} may
find.   Some of these values of $g$ and $h$
significantly exceed values extracted from experiment in Refs.~\cite{Ahmed:2001xc,Stewart:1998ke}
and Ref.~\cite{Mehen:2004uj}, respectively. They also exceed estimates based 
on the quark model~\cite{Becirevic:1999fr}, extraction from lattice QCD simulations
\cite{McNeile:2004rf,Abada:2003un} as well as 
sum rule constraints~\cite{Pirjol:1997nh}.
Again, however, it is not clear what can be concluded when comparing parameters which are by
themselves unphysical and whose definition depends upon the details of a calculation.  Of
perhaps more concern is that these fits produce large values for the
hyperfine coefficients. The operators which cause hyperfine splitting
should be $1/m_Q$ suppressed compared to the leading order ones.  Set
(d) is an example of a solution where $|g|$ is near $|g^\prime|$ and
$\Delta_S$ is within 30 MeV of $\Delta_H$. The relevance of that
result will become apparent in the next section.  Finally, an example
fit where $g$ and $h$ are restricted to lie between 0 and 1 yields
central values of parameters as follows: 
\bea
|g|=0.43,\, |g^\prime|=0,\,
h=0.31,\, \delta_H = 25 \,{\rm MeV}, \, \delta_S = 443 \,{\rm MeV},\, \qquad \qquad\nn \\
\Delta_H = 124 \,{\rm MeV},\, \Delta_S = 131 \,{\rm MeV}, \,
a_H = 2.4,\, a_S =-0.3,\, \Delta_H^{(a)}=-0.2,\, \Delta_S^{(a)}= 0.1 \, .\nn
\eea  
These parameter values lead to a prediction for the mass spectrum that also agrees with Eq.~(\ref{rm}).

The underconstrained nature of the various fits makes strong
conclusions impossible. In particular, the uncertainty in the parameter space is very large
and the uncertainty in individual parameters is much greater than indicated by the 
errors quoted in the individual fits listed above. The situation should improve with a global
fit to both masses and decay rates  which uses a consistent set of next-to-leading order calculations that 
include the excited states. This work is in progress.

\section{Hyperfine Splittings}

In this section we study the one loop corrections to the hyperfine splittings 
to see if HH$\chi$PT can provide insight into the observed near equality of the hyperfine splittings.
Using the formulae in the Appendix we find that the next-to-leading order difference between even-parity and 
odd-parity hyperfine splittings in the strange sector is given by
\bea\label{hypSH}
(m_{S^*_3}&&\!\!\! - m_{S_3})-(m_{H^*_3} - m_{H_3})  = (m^0_{S^*_3}-m^0_{S_3})-(m^0_{H^*_3}-m^0_{H_3}) \nn \\
&&+\frac{g^{\prime 2}}{f^2}\left[
\frac{2}{3} K_1(m^0_{S_1}-m^0_{S_3^*},m_K) + \frac{2}{9} K_1(m^0_{S_3}-m^0_{S_3^*},m_\eta) 
+ \frac{4}{3} K_1(m^0_{S^*_1}-m^0_{S^*_3},m_K) \right. \nonumber \\
&&\left. \qquad  + \frac{4}{9} K_1(0,m_\eta) - 2 K_1(m^0_{S_1^*}-m^0_{S_3},m_K) 
-\frac{2}{3}  K_1(m^0_{S_3^*}-m^0_{S_3},m_\eta) \right] \nonumber \\
&&-\frac{g^2}{f^2}\left[
\frac{2}{3} K_1(m^0_{H_1}-m^0_{H_3^*},m_K) + \frac{2}{9} K_1(m^0_{H_3}-m^0_{H_3^*},m_\eta) 
+ \frac{4}{3} K_1(m^0_{H^*_1}-m^0_{H^*_3},m_K) \right. \nonumber \\
&&\left. \qquad  + \frac{4}{9} K_1(0,m_\eta) - 2 K_1(m^0_{H_1^*}-m^0_{H_3},m_K) 
-\frac{2}{3}  K_1(m^0_{H_3^*}-m^0_{H_3},m_\eta) \right] \nonumber \\
&&+\frac{h^2}{f^2}\left[
2 K_2(m^0_{H_1^*}-m^0_{S_3^*},m_K)+ \frac{2}{3} K_2(m^0_{H_3^*}-m^0_{S_3^*},m_\eta)
-2 K_2(m^0_{H_1}-m^0_{S_3},m_K) \right. \nonumber \\
&&\left. \qquad - \frac{2}{3} K_2(m^0_{H_3}-m^0_{S_3},m_\eta)
-2 K_2(m^0_{S_1^*}-m^0_{H_3^*},m_K)- \frac{2}{3} K_2(m^0_{S_3^*}-m^0_{H_3^*},m_\eta)
\right. \nonumber \\
&&\left. \qquad  + 2 K_2(m^0_{S_1}-m^0_{H_3},m_K)+\frac{2}{3} K_2(m^0_{S_3}-m^0_{H_3},m_\eta) \right]  
\eea
Suppose one imposes at tree level the condition that all hyperfine splittings in each of the doublets are degenerate:
\bea 
m^0_{H^*_a} -m^0_{H_a} = m^0_{S^*_a} -m^0_{S_a} = \Delta 
\eea
This can be arranged by invoking the tree level prediction of the parity doubling model, $\Delta_H =\Delta_S =\Delta$ and neglecting the terms proportional
to $m_q$ in Eq.~(\ref{tree}), which are formally higher order in the power counting. 
Then $m^0_{S^*_a} - m^0_{H^*_a} = m^0_{S_a} - m^0_{H_a}$ and it is easy to verify
that all contributions proportional to $h^2$ vanish, and the remaining terms are:
\bea
(m_{S^*_3}&&\!\!\! -m_{S_3})-(m_{H^*_3}-m_{H_3})= \frac{{g^\prime}^2}{f^2}\left[\frac{2}{3} K_1(-\Delta,m_\pi) +\frac{2}{9} K_1(-\Delta,m_\eta)
+ \frac{16}{9} K_1(0,m_K) \nn \right. \\
&& \left.   -2 K_1(\Delta,m_K) - \frac{2}{3}K_1(\Delta,m_\eta) \right] - \frac{g^2}{f^2}\left[\frac{2}{3} K_1(-\Delta,m_\pi) 
+\frac{2}{9} K_1(-\Delta,m_\eta)\right. \nn \\
&&  \left. 
+ \frac{16}{9} K_1(0,m_K)    -2 K_1(\Delta,m_K)-\frac{2}{3}K_1(\Delta,m_\eta) \right]  
\eea
This vanishes if $g^2 = g^{\prime 2}$, which is consistent with the parity doubling model prediction. A similar cancellation
occurs for the nonstrange hyperfine splittings. So the parity doubling model explanation for the equality of the $j^p =\frac{1}{2}^-$
and $\frac{1}{2}^+$ hyperfine splittings is robust in the sense that one loop 
corrections do not spoil the prediction. 

The parity doubling model prediction for the axial couplings and hyperfine splittings singles out a
subspace of the parameter space of HH$\chi$PT that is preserved under renormalization group evolution. 
From our mass formulae it is easy to derive the following 
renormalization group equations for the renormalized parameters $\Delta_H$ and $\Delta_S$:
\bea 
\mu^2 \frac{d}{d \mu^2} \Delta_H &=&
\frac{4 g^2}{9 \pi^2 f^2}\Delta_H^3   \\
&& - \frac{h^2}{3 \pi^2 f^2} (\Delta_S - \Delta_H) \left[ 3 (\delta_S-\delta_H)^2
-\frac{3}{2}(\Delta_S - \Delta_H)(\delta_S-\delta_H) + \frac{7}{16} (\Delta_S-\Delta_H)^2  \right] \, . \nn \\
\mu^2 \frac{d}{d \mu^2} \Delta_S &=&
\frac{4 {g^\prime}^ 2}{9 \pi^2 f^2}\Delta_S^3   \\
&& + \frac{h^2}{3 \pi^2 f^2} (\Delta_S - \Delta_H) \left[ 3 (\delta_S-\delta_H)^2
-\frac{3}{2}(\Delta_S - \Delta_H)(\delta_S-\delta_H) + \frac{7}{16} (\Delta_S-\Delta_H)^2  \right] \, . \nn
\eea
which leads to 
\bea\label{rgespl}
\mu^2 \frac{d}{d \mu^2} (\Delta_S-\Delta_H) &=&\frac{4 }{9 \pi^2 f^2}({g^\prime}^2 \Delta_S^3 - g^2 \Delta_H^3) \\
&& \hspace{-0.6 in}+ \frac{2 h^2}{3 \pi^2 f^2} (\Delta_S - \Delta_H) \left[ 3 (\delta_S-\delta_H)^2
-\frac{3}{2}(\Delta_S - \Delta_H)(\delta_S-\delta_H)
 + \frac{7}{16} (\Delta_S-\Delta_H)^2  \right] \,. \nn
\eea
\begin{figure}[!t]
  \centerline{\epsfysize=7.0truecm \epsfbox[40 465 560 690]{gr.ps}  }
 {\tighten
\caption[1]{One-loop diagrams for renormalization of the coupling $g$.
$H$ fields are single lines, $S$ fields are double lines and Goldstone bosons are dotted lines. 
Diagrams for renormalization of the coupling $g^\prime$ are obtained by interchanging $H$ and $S$ fields.}
\label{gr} }
\end{figure}

We also derive the one loop renormalization group equation for the couplings $g$ and $g^\prime$. For this we need the wavefunction renormalization of the
fields $H$ and $S$, which is obtained from the graphs in Fig.~\ref{se}, and the one loop corrections to the axial couplings. The relevant graphs for the
renormalization of $g$ are shown in  Fig.~\ref{gr}, and the graphs for $g^\prime$ can be obtained from those in Fig.~\ref{gr} by interchanging $H$ and $S$
lines. Note that we only need the ultraviolet divergences of these graphs to obtain the renormalization group equation.  Furthermore, the counterterms for the
wavefunction renormalization and the axial couplings are defined to be independent of $m_q$ and $m_Q$.~\footnote{If the theory is not renormalized this way,
dependence on the underlying theory parameters $m_q$ and $m_Q$ would no longer be explicit in the chiral Lagrangian.} Ultraviolet divergences proportional
to $m_q$ and $1/m_Q$ are absorbed into higher order counterterms. For example, a  divergence proportional to $m_q$ in the one-loop correction to the axial
coupling of the $H$ fields should be renormalized by counterterms  with structures  like ${\rm Tr}[\overline H_a H_b \, \slash \hskip -.25cm A_{bc} \gamma_5]
m^\xi_{ca}$,  ${\rm Tr}[\overline H_a H_b \, \slash \hskip -.25cm A_{ba} \gamma_5] m^\xi_{cc}$, etc.  Therefore we can ignore Goldstone boson masses and
hyperfine splittings in computing the ultraviolet divergences, which greatly simplifies the calculation. The graphs in Figs.~\ref{se}a, \ref{se}c, \ref{gr}a,
and \ref{gr}e vanish in this limit because the integrals are scaleless. Graphs in Figs.~\ref{gr}c and \ref{gr}d do not contribute either. This is  because the
$H$-$S$-$\pi$ coupling  in Figs.~\ref{gr}c and \ref{gr}d gives a factor of $v\cdot k$, where $k^\mu$ is the four-momentum of the external Goldstone boson.
Ultraviolet divergences in Figs.~\ref{gr}c and \ref{gr}d are proportional  to $v\cdot k$ and are canceled by counterterms with an additional covariant
derivative acting on the fields $A^\mu_{ab}$, such as  ${\rm Tr}[\overline H_a H_b \, iv\cdot D_{bc} \, \slash \hskip -.25cm A_{ca} \gamma_5] $. Therefore, all
that is needed  to obtain the running of $g$ are the ultraviolet divergent parts of Fig.~\ref{se}b and \ref{gr}b in the limit where Goldstone bosons and
hyperfine splittings are neglected. The running of $g^\prime$ is obtained  from Fig.~\ref{se}d and the analog Fig.~\ref{gr}b  with $S$ and $H$ lines
interchanged. The result can  be obtained from the corresponding graphs for the renormalization of $g$ by simply substituting $g \leftrightarrow g^\prime$ and
$\delta_S - \delta_H \to -(\delta_S - \delta_H)$. The renormalization group equations for  $g$ and $g^\prime$ are
\bea\label{rge}
\mu \frac{d}{d \mu} g &=& -\frac{h^2}{4 \pi^2 f^2} (\delta_S -\delta_H)^2(g^\prime + 8 \,g) \nn\\
\mu \frac{d}{d \mu} g^\prime  &=& - \frac{h^2}{4 \pi^2 f^2}(\delta_H -\delta_S)^2(g  + 8\,g^\prime) \, ,
\eea
which can be rewritten as 
\bea\label{rgecoupling}
\mu \frac{d}{d \mu} (g + g^\prime) &=& - \frac{9 h^2}{4 \pi^2 f^2}(\delta_H -\delta_S)^2 (g + g^\prime) \nn\\
\mu \frac{d}{d \mu} (g - g^\prime)  &=& - \frac{7 h^2}{4 \pi^2 f^2}(\delta_H -\delta_S)^2 (g - g^\prime) \, .
\eea
To understand the significance of this result, consider the naive quark model prediction $g^\prime = g/3$~\cite{Falk:1992cx}.  From the renormalization group
equations in Eq.~(\ref{rge}) one sees that $g$ and $g^\prime$ vary with changes of the renormalization scale in such a way that the condition $g^\prime = g/3$
can only hold at one value of $\mu$. The quark model prediction is meaningless beyond tree level  without also specifying a particular renormalization scheme
and scale at which the relation is expected to hold. However, if $g = \pm g^\prime$ holds at any $\mu$, it will hold for all $\mu$ (at least at one loop
order). Also, if $g^2$ = $g^{\prime 2}$ and $\Delta_S = \Delta_H$ the right hand side of Eq.~(\ref{rgespl}) vanishes. Thus the predictions of the parity
doubling model, $\Delta_H=\Delta_S$  and $g  = - g^{\prime}$, are invariant under renormalization group flow in  HH$\chi$PT to one loop order.

\section{HQET and Predictions for Excited $B$ mesons}

In this section, we comment on the theoretical expectations for the
spectrum of excited even-parity bottom mesons which have yet to be
discovered.  Our HH$\chi$PT results for the charmed meson spectrum may
be used, but there are unknown $O(1/m_Q)$ effects which make it
difficult to obtain precise predictions for the $B$ meson. For finite
quark masses, to obtain the bottom meson spectrum from the charmed meson
results, the hyperfine operators should be rescaled by $m_c/m_b$,
which is not very well determined. Other parameters can receive
$O(\Lambda_{\rm QCD}/m_c - \Lambda_{\rm QCD}/m_b)$ corrections. For
instance, the reduced kinetic energy of the $b$ quark significantly
reduces the mass splitting of the $H$ and $S$ doublets in the $b$
sector relative to what is observed in the charmed system. These
$O(1/m_Q)$ corrections introduce significant uncertainty in HH$\chi$PT
predictions.

We will instead use the $O(1/m_Q)$ HQET formulae for the mass of a
heavy hadron $X$ which contains a heavy quark
$Q$~\cite{Manohar:2000dt}: \bea m^{(Q)}_X = m_Q + \bar \Lambda^X
-\frac{\lambda_1^X}{2 m_Q} +n_J \frac{\lambda_2^X}{2m_Q} \, , \eea
where $\lambda^X_{1}$ and $\lambda^X_{2}$ are hadronic matrix elements
of the HQET operators $\bar h (i D)^2 h$ and $g_s \bar h \sigma^{\mu
\nu} G_{\mu \nu} h/2$, respectively, and $n_J = +1$ for $J=1$ states
and $n_J = -3$ for $J=0$ states.  The first $1/m_Q$ correction,
$-\lambda_1^X/(2 m_Q)$, is the kinetic energy of the heavy quark. The
second $1/m_Q$ correction contributes to hyperfine splittings.  The
difference between the spin averaged masses of the $j^p=\frac{1}{2}^-$
and $j^p=\frac{1}{2}^+$ mesons, $\overline m^{(Q)}_H = (3
m^{(Q)}_{H^*} + m^{(Q)}_H)/4$ and $\overline m^{(Q)}_S = (3
m^{(Q)}_{S^*} + m^{(Q)}_S)/4$, respectively, is given by \bea
\overline m^{(Q)}_S -\overline m^{(Q)}_H = \bar \Lambda^S - \bar
\Lambda^H -\frac{\lambda_1^S}{2 m_Q} + \frac{\lambda_1^H}{2 m_Q} \, ,
\eea which leads to the following formulae for the splitting of the
even- and odd-parity states in the bottom sector: \bea \overline
m^{(b)}_S -\overline m^{(b)}_H = \overline m^{(c)}_S-\overline
m^{(c)}_H +(\lambda_1^S -\lambda_1^H)\left(\frac{1}{2 m_c} -
\frac{1}{2m_b}\right) \, .  \eea A recent global fit to $B$ decays
yields $\lambda_1^H = -0.20 \pm 0.06 \,{\rm
GeV}^2$~\cite{Bauer:2004ve}.  The parameter $\lambda_1^S$ is
unknown. From the spectroscopy of excited $j^p =\frac{3}{2}^+$ $D$ and
$B$ mesons, Ref.~\cite{Leibovich:1997em} extracts $\lambda_1^{3/2} -
\lambda_1^H = -0.23 \, {\rm GeV}^2$, where $\lambda_1^{3/2}$ is the
$\lambda_1$ matrix element for the $j^p=\frac{3}{2}^+$ doublet. The
sign here indicates that the kinetic energy of the heavy quark in the
excited heavy meson is larger than that in the ground state, which
agrees with intuition. We expect the kinetic energy of the heavy quark
in the $j^p =\frac{1}{2}^+$ states to be comparable to that of $j^p
=\frac{3}{2}^+$ states.  To estimate $\overline m^{(b)}_S$ with
conservative errors, we take $\lambda_1^S -\lambda_1^H = -0.2 \pm 0.1
\, {\rm GeV}^2$, $m_c =1.4$ GeV, and $m_b=4.8$ GeV to find
\bea\label{pre} \overline m^{(b)}_S -\overline m^{(b)}_H = \overline
m^{(c)}_S-\overline m^{(c)}_H - 50 \pm 25 \, {\rm MeV} \eea In the
bottom nonstrange sector, $m^{(b)}_{H_1} = 5279$ MeV and
$m^{(b)}_{H^*_1}= 5325$ MeV, which yields $\overline m^{(b)}_{H_1} =
5314$ and therefore Eq.~(\ref{pre}) predicts $\overline m^{(b)}_{S_1}
= 5696 \pm 30 \pm 25$ MeV. The first error comes from the uncertainty
in the charm nonstrange $j^p=\frac{1}{2}^+$ masses and the second is
the estimated uncertainty in $\lambda^S_1$. These states are well
above the threshold for $S$-wave pion decays to the ground state and
should be broad like their charm counterparts.

In the bottom strange sector, only the $0^-$ state with mass  $m^{(b)}_{H_3}= 5370$ MeV has been observed.
 To proceed we need to estimate the mass of the bottom strange $1^-$ state. Note that 
\bea\label{scale}
\frac{m_{H^*}^{(b)} - m_H^{(b)}}{ m_{H^*}^{(c)} - m_H^{(c)}} = \frac{m_{S^*}^{(b)} - m_S^{(b)}}{ m_{S^*}^{(c)} - m_S^{(c)}} 
=\frac{m_c}{m_b} \, .
\eea
up to $O(1/m_Q)$ corrections. Thus, all the hyperfine splittings in the bottom sector 
are related to those in the charm sector by a universal factor. Combining this with the measured 
value of $m_{H_1^*}^{(b)} - m_{H_1}^{(b)}$  leads to the prediction that $m_{H_3^*}^{(b)} - m_{H_3}^{(b)} 
= m_{S_3^*}^{(b)} - m_{S_3}^{(b)} = 46$ MeV, and $m_{S_1^*}^{(b)} - m_{S_1}^{(b)} = 42$ MeV. These predictions
have approximately 25\% uncertainty due to higher order $O(\Lambda_{\rm QCD}/m_c-\Lambda_{\rm   QCD}/m_b)$
corrections and the prediction for $m_{S_1^*}^{(b)} - m_{S_1}^{(b)} $ estimate has an additional 20\% uncertainty due to the poorly known $m_{S_1^*}^{(c)}$ and $m_{S_1}^{(c)}$  masses. Given these hyperfine splittings, one expects $\overline m^{(b)}_{H_3} = 5404$ MeV and then Eq.~(\ref{pre}) predicts $\overline m^{(b)}_{S_3} = 5702 \pm 25$ MeV. Here the error is dominated by our ignorance of $\lambda^1_S$. Note that the excited bottom strange
mesons are expected to lie well below the threshold for decays to ground state $B$ mesons and kaons
and should be narrow like $j^p=\frac{1}{2}^+$ charmed strange mesons.

\section{Conclusions}

We have enumerated the leading and subleading operators which
describe the even-parity charmed meson masses in heavy hadron chiral
perturbation theory (HH$\chi$PT).  We performed a loop
calculation to analyze the lowest lying even- and odd-parity charmed meson masses to ${\cal O}(Q^3)$.
There are nominally eleven unknown parameters in the prediction, and
only eight experimental masses.  Two of the parameters, the axial
coupling $g$ for the lowest doublet of charmed mesons, and the
coupling $h$ which dominates the strong decay between the even-parity
and ground state doublets, have been extracted from previous
calculations.  See Ref.~\cite{Stewart:1998ke} and
Ref.~\cite{Mehen:2004uj}, respectively.  However, the even-parity
states were not included in the extraction of $g$ in
Ref.~\cite{Stewart:1998ke}.  Also, the extraction of $h$ was only
performed at tree level.  Since these values for $g$ and $h$ were not
obtained under the same conditions as the mass calculations performed
in this paper, it is not clear that the values should be used in our
fit.  Indeed, if the values from Refs.~\cite{Stewart:1998ke} and
\cite{Mehen:2004uj} are used, it is not possible to obtain the
nonstrange even-parity masses as large as they are observed to be.  
If the $g$ and $h$ parameters are not fixed but simply constrained
to lie between 0 and 1, which is the prejudice from other analyses,
then a fit to the even-parity masses is possible. Because of the numerous
undetermined parameters, HH$\chi$PT can accomodate, but not explain, the unusual pattern of 
$SU(3)$ breaking observed in the excited charmed meson spectrum.
 
If we perform an unconstrained fit to the charmed meson mass spectrum
using all eleven parameters, many solutions are possible, including
ones whose values of $g$ and $h$ are not unreasonably far from their
previously extracted values.  However, then the parameter values found for the hyperfine operators  are of concern.  These hyperfine operators should have coefficients which scale as ${\cal O}(\Lambda^2_{QCD}/m_Q)$
relative to the ${\cal O}(\Lambda_{QCD})$ coefficients of the leading
operators.  The fact that unconstrained global fits find coefficients which are
sometimes larger for the hyperfine operators than for the leading
order operators may signal a problem in the $1/m_Q$ expansion. On the other hand, this may simply be a consequence of not properly incorporating the constraints on $g$ and $h$ from the decay widths. 
Before more definitive statements can be made, a global fit including even-parity intermediate states and terms up to ${\cal O}(Q^3)$ for both the odd-parity and even-parity meson decay rates must be done.  That will be the subject of a subsequent paper.

Next we consider the parity doubling model introduced in
Refs.~\cite{Bardeen:1993ae,Bardeen:2003kt,Nowak:1992um,Nowak:2003ra}. 
 While the parity doubling model is not
a result of QCD, but requires additional assumptions, it is
interesting because it provides an explanation of the observed
equality of the hyperfine splitting in the even-parity doublet and the
hyperfine splitting in the odd-parity doublet.  QCD symmetries alone
do not dictate any relationship between these hyperfine splittings.
While the parity doubling model provides an explanation for the
equality of the hyperfine splittings, the question we address here is
whether it is a {\sl natural} explanation.  That is, does it survive
beyond tree level?  Is it stable under RG flow?  We find that there
are ``fixed lines'' at $|g|=|g^\prime|$. (These are axial operator 
coefficients from Eq.~(\ref{axial}).) That is, if at any time in
their evolution $g=g^\prime$ or $g=-g^\prime$, RG analysis shows that
the relationship will be maintained.  This in turn assures that if at
tree level the parameters $\Delta_H$ and $\Delta_S$ in Eq.~(\ref{mass})
are equal, they remain so to one loop.  This lends credence to the
parity doubling model. The stability found in the parity doubling model 
does not exist for other models, such as the nonrelativistic quark model,
which predicts $g^\prime = g/3$. Going back to the parameter fit, we do find that solutions with $|g|$ near
$|g^\prime|$ are possible, as are fits with $\Delta_H$ near $\Delta_S$.
However, such fits yield values for $\Delta_H$ and $\Delta_S$ which
are larger than expected by power counting. In addition, there are fits which reproduce 
the observed hyperfine splittings without $|g|\approx |g^\prime|$.

Finally, we discuss how the charmed meson spectrum results can be used
to make predictions for the analog $B$ meson spectrum.  It is
necessary to know the charm and bottom quark masses in order to
rescale the operators, which brings in significant uncertainty.  Also,
there are additional $1/m_Q$ operators with unknown parameters.
However, it is possible to use heavy quark effective theory to
estimate that the even-parity strange spin-zero $B$ meson has mass $\sim$
5667 MeV while its spin-one partner has mass $\sim$ 5714 MeV.  This places them
below the threshold for decay to a kaon and the ground state $B$.
Therefore, we expect narrow  $B^*_s$ meson analogs to the narrow
$D^*_s$ excited mesons.

R.P.S. and T.M. are supported in part by DOE grant 
DE-FG02-96ER40945. T.M. is also supported in part by DOE grant DE-AC05-84ER40150.  
T.M. would  like to thank the Aspen Center for Physics and the KITP in Santa Barbara 
where parts of this work were completed.

\section{Appendix}

We express our results in terms of the functions
\begin{eqnarray}
K_1(\eta,M) &=&  {1 \over 16 \pi^2} \left[(- 2 \eta^3 + 3 M^2 \eta)\, \ln\left(\frac{M^2}{\mu^2}\right)
+ 2 \eta(\eta^2-M^2)\, F\left(\frac{\eta}{M}\right) + 4 \eta^3-5 \eta M^2 \right] \nn \\
K_2(\eta,M) &=&  {1 \over 16 \pi^2} \left[(- 2 \eta^3 +  M^2 \eta) \, \ln\left(\frac{M^2}{\mu^2}\right)
+ 2 \eta^3\,  F\left(\frac{\eta}{M}\right) + 4 \eta^3- \eta M^2 \right] 
\end{eqnarray}
where
\begin{eqnarray}
F(x)  &=&  2\frac{\sqrt{1-x^2}}{x}\left[\frac{\pi}{2} - {\rm Tan}^{-1}\left(\frac{x}{\sqrt{1-x^2}}\right) \right]
\qquad  |x| < 1 \\
&=&  -2\frac{\sqrt{x^2-1}}{x} \, \ln \left(x+\sqrt{x^2 - 1} \,\right)  
\qquad  \qquad \quad |x| > 1
\end{eqnarray}
The function $K_1(\eta,M)$ appears whenever the virtual heavy meson inside the loop
is in the same doublet as the external heavy meson, while $K_2(\eta,M)$ appears 
when the virtual heavy meson is from the opposite parity doublet.

In the limit $M << \eta$ these functions become 
\begin{eqnarray}
K_1(\eta,M) &=&  {1 \over 16 \pi^2} \left[-2 \,\eta^3 \,\ln\left(\frac{4\eta^2}{\mu^2}\right)
+ 3 \, \eta \,M^2 \,\ln\left(\frac{4\eta^2}{\mu^2}\right)  
+\frac{3\,M^4}{4\,\eta} \,\ln\left(\frac{M^2}{4\eta^2}\right)  + ...\right] \nn \\
K_2(\eta,M) &=&  {1 \over 16 \pi^2}  \left[-2 \eta^3 \,\ln\left(\frac{4\eta^2}{\mu^2}\right)
+  \eta\, M^2 \,\ln\left(\frac{4\eta^2}{\mu^2}\right) 
-\frac{M^4}{4\eta} \,\ln\left(\frac{M^2}{4\eta^2}\right)  + ...   \right] \, .
\end{eqnarray}
In these equations we have dropped polynomials of $\eta, M$. The functions $K_1(\eta,M)$ and $K_2(\eta,M)$ have well-defined $M \to
0$ limits. Furthermore, the dependence on $M$ is analytic when $M/\eta \to 0$, so in this  limit the $S$ fields can be integrated
out and their effect on the chiral corrections can be absorbed into local counterterms as expected. This limit is not relevant to
the real world as $\eta \sim M$.  In the opposite limit, $\eta =0$, which is relevant for loops in which external and virtual heavy
mesons are the same, 
\begin{eqnarray}
K_1(\eta,M)  &=& -\frac{M^3}{8 \pi} +{3 \over 16 \pi^2} \eta\, M^2 \,\ln\left(\frac{4\eta^2}{\mu^2}\right) +O(\eta^3) \nonumber  \\
K_2(\eta,M)  &=& {1 \over 16 \pi^2} \eta\, M^2 \,\ln\left(\frac{4\eta^2}{\mu^2}\right) +O(\eta^3)
\end{eqnarray}

Including the one loop diagrams we find:
\begin{eqnarray}
m_{H_1} &=& m^0_{H_1} +
{g^2 \over f^2}  \left[ {3 \over 2} K_1(m^0_{H_1^*}-m^0_{H_1},m_\pi) 
+ {1 \over 6} K_1 (m^0_{H_1^*}-m^0_{H_1},m_\eta) +
K_1( m^0_{H_3^*}-m^0_{H_1},m_K)\right] \nonumber\\ 
 && + {h^2 \over f^2} \left[ {3 \over 2} K_2(m^0_{S_1}-m^0_{H_1} ,m_\pi) \
+ {1 \over 6} K_2 (m^0_{S_1}-m^0_{H_1},m_\eta) +
K_2(m^0_{S_3}-m^0_{H_1},m_K)\right] \, .
\end{eqnarray}
\begin{eqnarray}
m_{H_3} &=& m^0_{H_3} +
{g^2 \over f^2}  \left[ 2 K_1(m^0_{H_1^*}-m^0_{H_3},m_K) 
+ {2 \over 3} K_1 (m^0_{H_3^*}-m^0_{H_3},m_\eta) \right]  \nn  \\ 
 && + {h^2 \over f^2} \left[ 2 K_2(m^0_{S_1}-m^0_{H_3},m_K) 
+ {2 \over 3} K_2 (m^0_{S_3}-m^0_{H_3},m_\eta)  \right] \, .
\end{eqnarray}
\begin{eqnarray}
m_{H_1^*} &=& m^0_{H_1^*} +
{g^2 \over f^2} \frac{1}{3} \left[ {3 \over 2} K_1(m^0_{H_1}-m^0_{H_1^*},m_\pi) 
+ {1 \over 6} K_1 (m^0_{H_1}-m^0_{H_1^*},m_\eta) +
K_1( m^0_{H_3}-m^0_{H_1^*},m_K)\right] \nonumber\\ 
&& + {g^2 \over f^2} \frac{2}{3} \left[ {3 \over 2} K_1(0,m_\pi) 
+ {1 \over 6} K_1 (0,m_\eta) +
K_1( m^0_{H_3^*}-m^0_{H_1^*},m_K)\right] \nonumber\\ 
&& + {h^2 \over f^2} \left[ {3 \over 2} K_2(m^0_{S_1^*}-m^0_{H_1^*} ,m_\pi) \
+ {1 \over 6} K_2 (m^0_{S_1^*}-m^0_{H_1^*},m_\eta) +
K_2(m^0_{S_3^*}-m^0_{H_1^*},m_K)\right] \, .
\end{eqnarray}
\begin{eqnarray}
m_{H_3^*} &=& m^0_{H_3^*} +
{g^2 \over f^2} \frac{1}{3} \left[ 2 K_1(m^0_{H_1}-m^0_{H_3^*},m_K) 
+ {2 \over 3} K_1 (m^0_{H_3}-m^0_{H_3^*},m_\eta) \right] \nonumber\\ 
&& + {g^2 \over f^2} \frac{2}{3} \left[ 2 K_1(m^0_{H_1^*}-m^0_{H_3^*},m_K) 
+ {2 \over 3} K_1 (0,m_\eta) \right] \nonumber\\ 
&& + {h^2 \over f^2} \left[ 2 K_2(m^0_{S_1^*}-m^0_{H_3^*} ,m_K) \
+ {2\over 3} K_2 (m^0_{S_3^*}-m^0_{H_3^*},m_\eta)  \right] \, . 
\end{eqnarray}
\begin{eqnarray}
m_{S_1} &=& m^0_{S_1} +
{{g^\prime}^2 \over f^2}  \left[ {3 \over 2} K_1(m^0_{S_1^*}-m^0_{S_1},m_\pi) 
+ {1 \over 6} K_1 (m^0_{S_1^*}-m^0_{S_1},m_\eta) +
K_1( m^0_{S_3^*}-m^0_{S_1},m_K)\right] \nonumber \\ 
 && + {h^2 \over f^2} \left[ {3 \over 2} K_2(m^0_{H_1}-m^0_{S_1} ,m_\pi) \
+ {1 \over 6} K_2 (m^0_{H_1}-m^0_{S_1},m_\eta) +
K_2(m^0_{H_3}-m^0_{S_1},m_K)\right]  \, .  
\end{eqnarray}
\begin{eqnarray}
m_{S_3} &=& m^0_{S_3} +
{{g^\prime}^2 \over f^2}  \left[ 2 K_1(m^0_{S_1^*}-m^0_{S_3},m_K) 
+ {2 \over 3} K_1 (m^0_{S_3^*}-m^0_{S_3},m_\eta) \right]  \nn \\ 
 && + {h^2 \over f^2} \left[ 2 K_2(m^0_{H_1}-m^0_{S_3},m_K) 
+ {2 \over 3} K_2 (m^0_{H_3}-m^0_{S_3},m_\eta)  \right]  \, .  
\end{eqnarray}
\begin{eqnarray}
m_{S_1^*} &=& m^0_{S_1^*} +
{{g^\prime}^2 \over f^2} \frac{1}{3} \left[ {3 \over 2} K_1(m^0_{S_1}-m^0_{S_1^*},m_\pi) 
+ {1 \over 6} K_1 (m^0_{S_1}-m^0_{S_1^*},m_\eta) +
K_1( m^0_{S_3}-m^0_{S_1^*},m_K)\right] \nonumber\\ 
&& + {{g^\prime}^2 \over f^2} \frac{2}{3} \left[ {3 \over 2} K_1(0,m_\pi) 
+ {1 \over 6} K_1 (0,m_\eta) + K_1( m^0_{S_3^*}-m^0_{S_1^*},m_K)\right] \nonumber\\ 
&& + {h^2 \over f^2} \left[ {3 \over 2} K_2(m^0_{H_1^*}-m^0_{S_1^*} ,m_\pi) \
+ {1 \over 6} K_2 (m^0_{H_1^*}-m^0_{S_1^*},m_\eta) +
K_2(m^0_{H_3^*}-m^0_{S_1^*},m_K)\right]   \, . 
\end{eqnarray}
\begin{eqnarray}
m_{S_3^*} &=& m^0_{S_3^*} +
{{g^\prime}^2 \over f^2} \frac{1}{3} \left[ 2 K_1(m^0_{S_1}-m^0_{S_3^*},m_K) 
+ {2 \over 3} K_1 (m^0_{S_3}-m^0_{S_3^*},m_\eta) \right] \nonumber\\ 
&& + {{g^\prime}^2 \over f^2} \frac{2}{3} \left[ 2 K_1(m^0_{S_1^*}-m^0_{S_3^*},m_K) 
+ {2 \over 3} K_1 (0,m_\eta) \right] \nonumber\\ 
&& + {h^2 \over f^2} \left[ 2 K_2(m^0_{H_1^*}-m^0_{S_3^*} ,m_K) 
+ {2\over 3} K_2 (m^0_{H_3^*}-m^0_{S_3^*},m_\eta)  \right]  \, . 
\end{eqnarray}

We agree with Ref.~\cite{Jenkins:1992hx} for the $H$  fields in
the limit where $m_\pi \rightarrow 0$, $m_\eta^2 \rightarrow {4 \over 3} m_K^2$ and
 $\eta/M \ll 1$. Our answer also agrees with that of 
Ref.~\cite{Becirevic:2004uv}, which computes mass corrections to the $H$ and $S$ masses 
including $SU(3)$ breaking corrections but not hyperfine splittings.

\end{document}